\documentclass[manuscript]{aastex61}

\accepted{\today}
\def\MO{M_\odot}
\def\ME{M_\oplus}
\shorttitle{First sub-arcsecond submillimeter-wave [C I] image of 49 Ceti with ALMA}
\shortauthors{Higuchi et al.}

\begin{document}
\title{First sub-arcsecond submillimeter-wave [C I] image of 49 Ceti with ALMA}

\correspondingauthor{Aya E. Higuchi}
\email{aya.higuchi@nao.ac.jp}

\author[0000-0002-9221-2910]{Aya E. Higuchi}
\affil{National Astronomical Observatory of Japan, Osawa, Mitaka, Tokyo 181-8588, Japan}

\author{Kazuya Saigo}
\affiliation{National Astronomical Observatory of Japan, Osawa, Mitaka, Tokyo 181-8588, Japan}

\author{Hiroshi Kobayashi}
\affiliation{Department of Physics, Nagoya University, Furo-cho, Chikusa-ku, Nagoya, Aichi, 464-8602, Japan}

\author{Kazunari Iwasaki}
\affiliation{National Astronomical Observatory of Japan, Osawa, Mitaka, Tokyo 181-8588, Japan}

\author{Munetake Momose}
\affiliation{College of Science, Ibaraki University, Bunkyo 2-1-1, Mito 310-8512, Japan}

\author{Kang Lou Soon}
\affiliation{College of Science, Ibaraki University, Bunkyo 2-1-1, Mito 310-8512, Japan}

\author{Nami Sakai}
\affiliation{RIKEN Cluster for Pioneering Research, 2-1, Hirosawa, Wako-shi, Saitama 351-0198, Japan}

\author{Masanobu Kunitomo}
\affiliation{Department of Physics, School of Medicine, Kurume University, 67 Asahi-machi, Kurume, Fukuoka 830-0011, Japan}

\author{Daisuke Ishihara}
\affiliation{Institute of Space and Astronautical Science, Japan Aerospace Exploration Agency, 3-1-1 Yoshinodai, Chuo-ku, Sagamihara, Kanagawa 252-5210, Japan}

\author{Satoshi Yamamoto}
\affiliation{Department of Physics, The University of Tokyo, Hongo, Bunkyo-ku, Tokyo 113-0033, Japan}

\begin{abstract}

We present the first sub-arcsecond images of 49 Ceti in the [C I] $^{3}${\it{P}}$_{1}$--$^{3}${\it{P}}$_{0}$ emission 
and the 614~$\micron$ dust continuum emission observed with ALMA, as well as that in the CO($J$=3--2) emission 
prepared by using the ALMA archival data.
The spatial distribution of the 614~$\micron$ dust continuum emission is found to have a broad-ring 
structure with a radius of about 100~au around the central star.
A substantial amount of gas is also associated with 49 Ceti.
The [C I] emission map shows two peaks inside the dust ring, and its overall extent is comparable to that of the 
dust continuum emission and the CO emission.
We find that the [C I]/CO($J$=3--2) intensity ratio significantly varies along the major axis. 
The ratio takes the minimum value of 1.8 around the dust peak position, and increases inwards and outwards. 
The enhanced ratio around the central star ($\sim$3) likely originates from the stellar UV radiation, 
while that in the outer disk ($\sim$10) from the interstellar UV radiation. 
Such complex distributions of the [C I] and CO($J$=3--2) emission will be a key to understand the origin of the gas in 49 Ceti, 
and will also provide a stringent constraint on physical and chemical models of gaseous debris disks.  

\end{abstract}
\section{Introduction} \label{sec:intro}

Recently, several debris disks harboring a gas component have been discovered in survey observations at optical, infrared, 
and radio wavelengths,
and its origin has been discussed in terms of the evolution of protoplanetary disks and the formation of planetary bodies.
In fact, many debris disks are known to reveal submillimeter-wave CO emission, 
e.g., 49~Ceti \citep{hug17}, $\beta$ Pictoris \citep{den14}, and 15 others or more \cite[e.g.,][]{kos13, moo13, moo17, mat17, hug18}.
In addition to the CO emission, the submillimeter-wave [C I] emission has been observed toward a few debris disks.
For instance, \cite{hig17} detected the bright [C I] $^{3}${\it{P}}$_{1}$--$^{3}${\it{P}}$_{0}$ toward 
49 Ceti and $\beta$ Pictoris with the ASTE 10~m telescope.
The [C I] emission has also been observed with ALMA toward $\beta$ Pictoris \citep{cat18} and HD~131835 \citep{kra18}.
Interestingly, the [C I] intensity is found to be higher than the CO intensity, which contrasts with the protoplanetary 
disk case \citep{tsu15, kam16}.
However, the origin of the strong [C I] emission from debris disks is not well understood.

49 Ceti is a famous A-type star, located at d = 57$\pm$0.3~pc \citep{gai18}, at the age of 12 -- 50~Myr \citep{zuc12, zuc19}. 
The CO emission was first detected with JCMT \cite[e.g.,][]{zuc95, den05}, partially resolved with SMA \citep{hug08}, 
and clearly imaged with ALMA \citep{hug17, moo19}.
Thanks to the high-resolution observation with ALMA (0\farcs56$\times$0\farcs45), \cite{hug17} 
found tentative evidence for a warp in the CO emission.
According to the ASTE observation by \cite{hig17}, the [C I] line profile resembles that of CO.
This result suggests that atomic carbon (C) would coexist with CO in the debris disks, and C is likely formed 
by the photodissociation of CO.
Furthermore, a simple chemical model indicates a small number of H$_{2}$ molecules in the gas disk.
However, the above interpretation depends on the physical parameters, such as visual extinction and dust size, 
employed in chemical models, and thus it is necessary to resolve 
the spatial distribution of the disk in order to understand the effect of UV radiation from the central star and 
from interstellar space.
Although the [C I] emission from $\beta$ Pictoris \citep{cat18} and HD~131835 \citep{kra18} has been imaged with ALMA, 
a detailed comparison between their CO and [C I] images is difficult due to an insufficient signal to noise ratio of 
the [C I] data. 
In this study, we present a high-quality image of 49 Ceti in the [C I] emission observed with ALMA, 
and report the spatial structure of the [C I] emission at a resolution of 0\farcs5 (30~au).

%with the velocity resolution of 0.1~km~s$^{-1}$.

\section{Observations and Imaging}

49 Ceti was observed with ALMA during Cycle 5 in its compact configuration by using the Band 8 receivers.
The observing condition was excellent, where the precipitable water vapor (PWV) during the observation ranged from 0.19 to 0.81~mm.
The observation was done in 7 execution blocks with 43~12-m antennas.
The minimum and maximum baseline lengths achieved during the observation were 15~m and 500~m, respectively.
On-source integration time was 5.9~hours.
The synthesized beam for the 614~$\micron$ dust continuum is 0\farcs51$\times$0\farcs43 with P.A.=  $-$ 84.8$^{\circ}$, while 
that for the \lbrack{C I}\rbrack~$^{3}${\it{P}}$_{1}$--$^{3}${\it{P}}$_{0}$ line (492.161~GHz) is 0\farcs50$\times$0\farcs42 
with P.A.= $-$ 84.8$^{\circ}$.
%{\bf The synthesized beam for CO line in the archival data is 0\farcs56$\times$0\farcs45 with P.A.= - 87.2$^{\circ}$.}
At the distance of 49 Ceti (d$\sim$57~pc), the angular size of 0\farcs5 corresponds to 30~au.
In this observation, the atmospheric water line at 183~GHz was simultaneously observed with water vapor radiometers 
to measure a temporal variation of the water vapor column in each antenna beam, which was used to reduce the atmospheric phase noise. 
The quasars J2258-2758, J0141-0928, J2253+1608. and J0423-0120 were employed to calibrate amplitude, bandpass, and complex gain fluctuations, 
depending on the observing schedule.

In addition, we employed the ALMA archival data of the CO($J$=3--2) line emission observed toward this source.
The synthesized beam for the CO line is 0\farcs56$\times$0\farcs45 with P.A.= $-$ 87.2$^{\circ}$, which is
comparable to those for the 614~$\micron$ continuum and the [C I] line.
Detailed information of the CO observation is described in \cite{hug17}.

Data reduction was performed by using version 5.3.0 of the Common Astronomy Software Applications (CASA) package \citep{mcm07}.
The 614~$\micron$ continuum map was obtained from a combination of all the line-free channels and by using natural weighting.
[C I] and CO images were obtained by using natural weighting.
The CASA task $\it{tclean}$ was used to Fourier-transform the visibility data and to deconvolve the dirty 
images at a velocity interval of 0.10~km~s$^{-1}$.
For the [C I] and CO images, the same parameters in $\it{tclean}$ were used.
In this paper, we present results of the 614~$\micron$ dust continuum, the [C I] $^{3}${\it{P}}$_{1}$--$^{3}${\it{P}}$_{0}$ line emission,
and the CO($J$=3--2) line emission.

\section{Results and Discussions}

\subsection{614~$\micron$ continuum emission}

Figure \ref{fig1}(a) shows the 614~$\micron$ dust continuum map at a 0\farcs5 resolution. 
The peak flux is obtained to be 1.1$\pm$0.1~mJy~beam$^{-1}$, while the integrated flux to be 24.4$\pm$0.1~mJy 
(above the 5~$\sigma$ of the observed image) by using Polyline drawing tool in CASA viewer.
The 614~$\micron$ continuum emission is distributed to a broad-ring structure around the central star.
The overall extent of the emission is estimated by the broad-ring model reported by \cite{hug17}.
Details of the model are described in Appendix A.
We derive the inner and outer radii of the distribution to be 60$\pm$10~au and 250$\pm$20~au, respectively (see Table 1).
%is 325~au and 65~au along the major and minor axes, respectively.
%above the 3σ emission of observed images

Although the 850~$\micron$ continuum emission also reveals a similar structure \citep{hug17}, 
the broad-ring shape is obtained with a higher signal to noise ratio (S/N) in our 614~$\micron$ continuum observation.
The dust temperature derived from the model fitting of the spectral energy distribution is 57~K \citep{hol17}, 
while the mean excitation temperature is reported to be 32~K \citep{hug17}.
Hence, we assume the range of dust temperature to be from 30 to 100~K based on these reports.
Assuming optically-thin thermal dust emission and a dust mass opacity of 
$\kappa_{\nu}$=2.3$\times$$(\nu/230~\rm{GHz})^{\beta}$~cm$^{2}$g$^{-1}$ with $\beta$ of 0.7 \citep{moo17}, 
we estimate the total dust mass to be less than 0.15~$\ME$.

\subsection{[C I] $^{3}${\it{P}}$_{1}$--$^{3}${\it{P}}$_{0}$ and CO($J$=3--2) line emission}

\begin{figure}
\epsscale{1.2}
\plotone{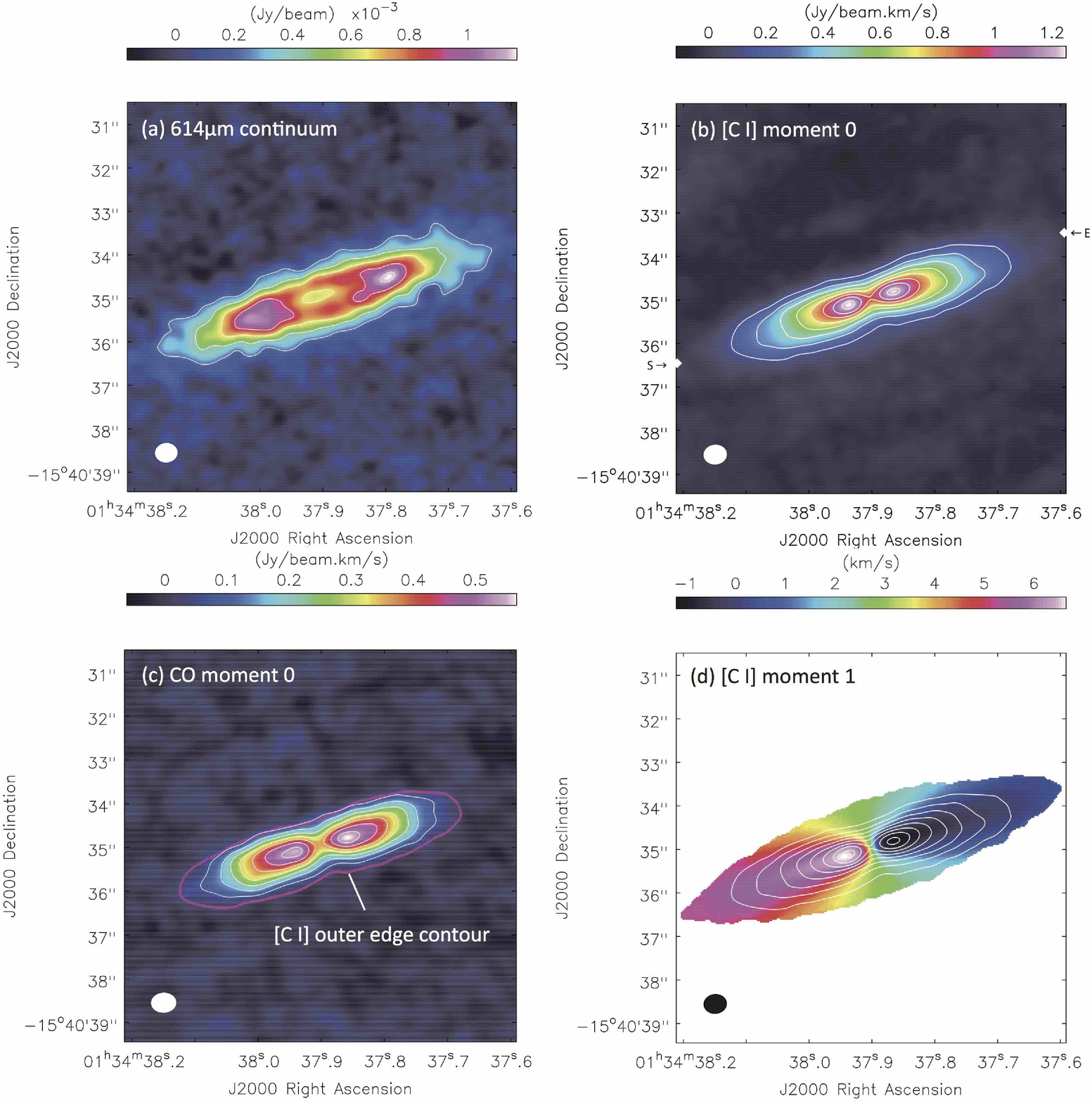}
\caption{
(a) The 614~$\micron$ continuum image. Contours are 5$\sigma$, 10$\sigma$, 15$\sigma$, 20$\sigma$, 25$\sigma$ levels 
(1$\sigma$ = 0.045~mJy~beam$^{-1}$). 
(b) The [C I] integrated intensity map (Velocity range = $-$6 to 11.5~km~s$^{-1}$).
Contours are 5$\sigma$, 10$\sigma$, 15$\sigma$, 20$\sigma$, 25$\sigma$, 30$\sigma$, 35$\sigma$, 40$\sigma$ levels 
(1$\sigma$ = 30~mJy~beam$^{-1}$~km~s$^{-1}$).
The starting (S) and end (E) positions of the major axis are indicated.
(c) The CO integrated intensity map (Velocity range = $-$6 to 11.5~km~s$^{-1}$). 
Contours are 5$\sigma$, 10$\sigma$, 15$\sigma$, 20$\sigma$, 25$\sigma$, 30$\sigma$, 35$\sigma$, 40$\sigma$ levels (1$\sigma$ = 15~mJy~beam$^{-1}$~km~s$^{-1}$).
The pink color contour shows the 5$\sigma$ level of the [C I] integrated intensity map.
(d) The [C I] velocity field map. Contours show the [C I] integrated intensity with the same contour levels of (b).}
\label{fig1}
\end{figure}

Figure \ref{fig1}(b) shows the integrated intensity map of the [C I] emission, 
where the velocity range for integration is from $-$6 to 11.5~km~s$^{-1}$.
%The extent of the [C I] emission is estimated by using the model shown in the Appendix.
The peak flux is obtained to be 0.52$\pm$0.03~Jy~beam$^{-1}$~km~s$^{-1}$, while the integrated flux to be 15.3$\pm$1.0~Jy~km~s$^{-1}$ 
(above the 5~$\sigma$ of the observed image) by using Polyline drawing tool in CASA viewer.
The [C I] disk size is measured by using the same model used for the dust continuum (Appendix A).
The inner and outer radii of the [C I] distribution are estimated to be 30$\pm$5~au and 195$\pm$20~au, respectively (Table 1).
The [C I] distribution has a double-peaked structure inside the broad ring distribution of the dust continuum emission.

Figure \ref{fig1}(c) shows the integrated intensity map of the CO ($J$=3--2) emission, 
where the velocity range for integration is from $-$6 to 11.5~km~s$^{-1}$. 
The peak flux is obtained to be 1.1$\pm$0.1~Jy~beam$^{-1}$~km~s$^{-1}$, while the integrated flux to be 5.7$\pm$0.3~Jy~km~s$^{-1}$ 
(above the 5~$\sigma$ of the observed image) by using Polyline drawing tool in CASA viewer.
The CO distribution also reveals a double-peaked feature inside the dust continuum peak, as in the case of the [C I] emission.
However, the separation of the two peaks is slightly wider than that for the [C I] case.
An overall extent of the CO emission is similar to that of the [C I] emission.

The velocity field map and the position-velocity (P-V) diagram of the [C I] emission (Figure \ref{fig1}(d) and Figure \ref{fig4}), 
clearly show the rotating disk as in the CO($J$=3--2) case, where the starting position of the P-V diagram is 
($\alpha_{J2000}$, $\delta_{J2000}$) = 
(01$^{\mbox h}$34$^{\mbox m}$38$^{\mbox s}.$210, ~$-$~$\!\!$15$^{\circ}$40$^{\arcmin}$36$\arcsec.$450),
and the end position is ($\alpha_{J2000}$, $\delta_{J2000}$) = 
(01$^{\mbox h}$34$^{\mbox m}$37$^{\mbox s}.$594, ~$-$~$\!\!$15$^{\circ}$40$^{\arcmin}$33$\arcsec.$450)
with P.A.= $-$ 72$^{\circ}$. The starting and end positions are indicated in Figure \ref{fig1}(b).
To prepare the P-V diagram, the CASA task {\it impv} is used.

The overall similarity of the distributions between the CO and [C I] emission suggests that 
the origin of C is photodissociation of CO by the UV radiation.
It should be noted that the total flux (spatially and spectrally integrated line flux) of the [C I] emission is 
higher than that of the CO ($J$=3--2) emission, which is consistent with the single dish result by \cite{hig17}.

\subsection{Intensity distribution}

In this observation, the [C I] emission from 49 Ceti has been spatially resolved for the first time with ALMA Band 8 observations.
Figure \ref{fig2} shows the intensity profiles of the [C I], CO, and continuum emission along the major axis (P.A.= $-$ 72$^{\circ}$) 
of the integrated intensity map of [C I] (Figure \ref{fig1}(b)).
The position axis, $x$, represents the projected distance from the central star.
The coordinates of the starting position is ($\alpha_{J2000}$, $\delta_{J2000}$) = 
(01$^{\mbox h}$34$^{\mbox m}$38$^{\mbox s}.$210, ~$-$~$\!\!$15$^{\circ}$40$^{\arcmin}$36$\arcsec.$450),
while those of the end position is ($\alpha_{J2000}$, $\delta_{J2000}$) = 
(01$^{\mbox h}$34$^{\mbox m}$37$^{\mbox s}.$594, ~$-$~$\!\!$15$^{\circ}$40$^{\arcmin}$33$\arcsec.$450).
%%%%%%%
%%%%%%%
They are the same as those for the P-V diagram (Figure \ref{fig4}).
The [C I] and CO profiles have a double peak inside the broad-ring structure of the dust continuum emission, 
as shown in Figure \ref{fig1}.
%%%
\footnote{The conversion equation from Jy~beam$^{-1}$ to K is given as :
$\left(\frac{T_{\rm{B}}}{\rm{K}}\right) = 1.222\times10^{6}
\left(\frac{\theta_{\rm{maj}}}{\rm{arcsec}}\right)^{-1}
\left(\frac{\theta_{\rm{min}}}{\rm{arcsec}}\right)^{-1}
\left(\frac{\nu}{\rm{GHz}}\right)^{-2}
\left(\frac{S}{\rm{Jy~beam^{-1}}}\right)$,
where $\theta_{\rm{maj}}$ is the major axis, $\theta_{\rm{min}}$ is the minor axis, 
$\nu$ is the frequency, and $S$ is the flux density.}

The [C I] and CO intensities are almost comparable to each other between the offset positions of 
70 and 140~au for both sides of the central star position.
On the other hand, the [C I] intensity is higher than the CO intensity inside and outside of the above range 
(i.e., $|x|$ $<$ 70~au and $|x|$ $>$ 140~au).
Thus, we can divide the distribution into the following three parts: the outer region with $|x|$ $>$ 140~au, 
the inner region with $|x|$ $<$ 70~au, and the intermediate region from 70 to 140~au.
The intermediate region just involves the peak of the dust continuum emission.
The radial distribution of the dust continuum emission is different from those gas components.

%and gas components are clearly different.

\subsection{Interpretation of the [C I]/CO intensity ratio}

Figure \ref{fig3} depicts the [C I]/CO intensity ratio along the major axis. 
As expected, the ratio has the minimum value around the intermediate region where the dust continuum emission peaks,
and increases inwards and outwards.
In order to interpret this characteristic variation of the [C I]/CO intensity ratio, we need to evaluate the column densities of C and 
CO at each offset position from the central star. However, their derivation is not straightforward. 
Recently, \cite{moo19} detected the $^{13}$CO($J$=2--1) emission as well as the $^{12}$CO($J$=2--1) emission from 49 Ceti by 
ALMA ACA (Atacama Compact Array) observation at a resolution of 6$\arcsec$.
According to their result, the flux ratio ($r$) of $^{12}$CO($J$=2--1)/$^{13}$CO($J$=2--1) is 2.3$\pm$0.2. 
Thus, the optical depth of the $^{12}$CO($J$=2--1) emission ($\tau$) is derived to as high as be 43$\pm$7 
by using the following relation:
%%%
\begin{equation}
r = \frac{1-\exp({-\tau})}{1-\exp({-\tau/77})},
\end{equation}
where the elemental $^{12}$C/$^{13}$C ratio of 77 \citep{wil94} is assumed and the slightly 
frequency difference between the $^{12}$CO and $^{13}$CO lines is ignored.
In addition, it is assumed that the two CO isotopologues have the same excitation temperature 
and are spectro-spatially co-located.
The $^{12}$CO($J$=2--1) line is indeed optically thick.
Likewise the CO($J$=3--2) line is most likely optically thick as well.

For this reason, the CO mass evaluated from the optically thin $^{13}$CO emission ($>$ 0.01~$\ME$) 
is higher by two orders of magnitude than the previous estimate ($>$ 10$^{-4}$~$\ME$) \citep{hug17,hig17}. 
Thus, the CO mass in 49 Ceti is higher by three orders of magnitude than the CO mass (3.4 $\times$ 10$^{-5}$~$\ME$) 
in $\beta$ Pictoris \citep{den14, mat17}.

We face a similar but more serious situation for the optical depth of the [C I] emission. 
Considering that the [C I] intensity is comparable to or even higher than the CO intensity, 
we cannot simply assume the optically thin condition 
for the [C I] emission. Thus, it is difficult to derive the column densities of CO and C from 
the observed intensities based on the current dataset. 
Therefore, we qualitatively discuss the behavior of the characteristic variation of the [C I]/CO intensity ratio in this study.

As shown in Figure \ref{fig3}, the [C I]/CO intensity ratio increases from the intermediate part to the inner and outer parts.
%toward the CO and [C I] peak positions. 
This trend can be interpreted as the different variation of the number density of C and CO along the major axis, 
the different variation of the excitation temperature of the [C I] and CO emission, or the mixture of both.

First, we discuss the case that the both [C I] and CO lines are optically thick and 
the variation of the [C I]/CO intensity ratio reflects the different variation of the excitation temperature.
Here, we simply assume the configuration shown in Figure \ref{fig4}.
We consider the stellar UV radiation and the interstellar UV radiation for a heating source.
%%%%
The stellar UV radiation will be attenuated by a large amount of C and CO in the mid-plane, as shown in Figure \ref{fig4}, 
and hence, their excitation temperature would monotonously decrease with increasing distance from the central star.
If a decreasing rate as a function of the radius is higher for CO than for [C I], the radially increasing feature of the [C I]/CO intensity ratio 
from the intermediate part to the outer part could be explained.
Since CO is expected to reside mainly in the dense region near the mid-plane, its cooling should be efficient.
Thus, the higher decreasing rate for CO seems reasonable.

In the inner part, we need to consider an additional effect, because the [C I]/CO intensity ratio increases toward the central star.
This means that the excitation temperature of [C I] becomes higher than that for CO as approaching the central star.
Since C can reside in the surface layer of the inner edge of the broad ring structure facing the central star (Figure \ref{fig4}), 
it is subject to stronger heating than CO.
Indeed, the [C I] emission peak is closer to the central star in the inner part than the CO emission peak.
Thus, the increasing [C I]/CO intensity ratio toward the central star seems reasonable with this picture.
In any case, the UV radiation from the central star and/or the interstellar UV radiation
plays an important role in the variation of the [C I]/CO intensity ratio.

%as in the case of the inner part. If the [C I] emission comes from the surface layer of the gas disk rather than 
%the CO emission as in the
%case of the protoplanetary disks, the excitation temperature of the [C I] emission would be higher 
%than that of the CO emission. 
%This picture seems reasonable, because C is a photodissociation product of CO by 

Second, we discuss the possibility that the variation of the C and CO abundances along the major axis contributes to the variation of 
[C I]/CO intensity ratios.
Even if the [C I] and CO emission is optically thick for the intermediate part, it could be optically thin in the inner and outer parts, 
where the gas density is likely lower than in the intermediate region.
Indeed, the brightness temperatures of [C I] and CO are about one-tenth of the individual peak intensities at the outer region,
implying a smaller amount of gas component in the outer region. 
Thus, the optical depths of the [C I] and CO emission there would be lower than those at the intermediate region. 
If so, the enhanced abundance of C due to the interstellar UV irradiation would contribute to the trend to some extent.
If we simply assume the optically thin condition for the both lines, the C/CO abundance ratio in the outer region is roughly 
estimated to be 100 by using the equations (1) and (2) presented by \cite{hig17}.
The situation near the central star may be considered in the same way. 
There is a dip of the brightness temperature near the central star, implying a smaller amount of C and CO compared with those 
in the intermediate part. 
This situation is also supported by the velocity structure in the position-velocity diagram (Figure \ref{fig4}).
If the optically thin condition were the case in the inner region, the C/CO abundance ratio would be esimated to be 50.

Above all, the combination of the both effects, the variation of the excitation temperature and that of the C and CO abundances,
would be the case for the inner and outer regions.
The C/CO abundance ratio in the gas is mainly determined by the strength of the UV radiation and the amount 
of H$_{2}$ molecules \cite[e.g.,][]{hig17, iwa19}.
The former controls the dissociation of CO to produce C and C$^{+}$, while the latter contributes to the reproduction of CO from C$^{+}$.
However, the optical depth problem hampers quantitative discussions on the C/CO abundance ratio.
For detailed modeling, we need spatially-resolved images of the other transitions of $^{12}$CO, $^{13}$CO and [C I], 
which allow us to make a correction for the optical depth effect.

\subsection{Future perspective for understanding origin of gas}

So far the origin of the gas in debris disks has been discussed on the basis of two scenarios: the secondary gas coming out from dust grains
and the primordial gas remaining from the protoplanetary stages.
Recently, \cite{kra18} predict that the spatial distributions of the [C I], CO and continuum emission can be different, according to 
their dynamical/chemical model for HD~131835 based on the secondary gas scenario. 
In their model, CO is outgassing from the planetesimal belt, i.e., the dust ring, and is photodissociated to form C. 
The gas will then spread viscously inwards and outwards.
Although there is no definitive evidence for outward viscous spreading in our data,
this picture (Figure 4 of \cite{kra18}) is qualitatively similar to our observational results (Figure \ref{fig2}).
They also point out a sharp drop of CO, which can be an evidence of the secondary gas scenario.
However, it is difficult to identify such a feature in the current dataset.
%Further detailed modeling is awaited for quantitative comparison.

If the gas is primordial, i.e., the remnant gas from the protoplanetary disk, 
the gas dispersal from the protoplanetary disk stage has not been completed yet in this object.
However, observational studies on dispersal mechanism are sparse, 
although it has been discussed for the cluster forming regions with intense 
UV radiation from nearby massive stars, e.g., Orion nebula \citep{ada04, arm11, man14}.
In these sources, photoevaporation plays a crucial role.
It is thus important to explore how this process can work in rather isolated sources like 49 Ceti with the reasonable time scale.

Broad-ring distributions of the dust continuum emission have been discovered in the continuum emission by recent 
ALMA observations toward 
several transitional disks such as DM~Tau \citep{kud18}, J1604--2130 \citep{don17, may18}, Sz~91 \citep{tsu19} and HD~135344B \citep{van19}.
For these sources, the CO emission associated with the disk is distributed inward of the dust ring.
In the inner region (dust hole), the stellar UV radiation may enhance photodissociation of CO to form C.
If so, the strong [C I] emission may be extended in the dust hole in the transitional disks.

Above all, the origin of the gas is still controversial, but the spatial distributions of the [C I], CO and continuum emission will be 
an important constraint to solve this problem.
One way to understand the origin of gas is to trace the gas dispersal stage from protoplanetary disks into debris disks.
Although it is challenging to carry out a long-time integration survey of the [C I] emission toward transitional disks, 
a statistical study of the spatial resolved [C I]/CO intensity ratios from protoplanetary disks to debris disks 
is highly awaited to trace the gas dispersal stage.

\bigskip
\acknowledgments
We thank the referee for the thoughtful and constructive comments.
The authors thank Gianni Cataldi for providing comments on a draft of the manuscript.
We also thank Attila Mo{\'o}r for providing results of their ACA observations prior to the publication.
We thank Aki Sato for her great contribution to this study.
We acknowledge the ALMA staff for the operation and maintenance of the observational instruments. 
This paper makes use of the following ALMA data:ADS/JAO.ALMA$\#$2017.0.00467.S and ADS/JAO.ALMA$\#$2012.1.00195.S.
ALMA is a partnership of ESO (representing its member states), NSF (USA) and NINS (Japan), together with NRC (Canada), NSC and ASIAA (Taiwan), 
and KASI (Republic of Korea), in cooperation with the Republic of Chile. 
The Joint ALMA Observatory is operated by ESO, AUI/NRAO and NAOJ.
This study is supported by KAKENHI (18H05222 and 18K03713).
Data analyses were carried out on a common-use data analysis computer system at the Astronomy Data Center, 
ADC, of the National Astronomical Observatory of Japan.

\facility{ALMA}
\software{CASA 5.3.0}

\begin{figure}
\epsscale{0.8}
\plotone{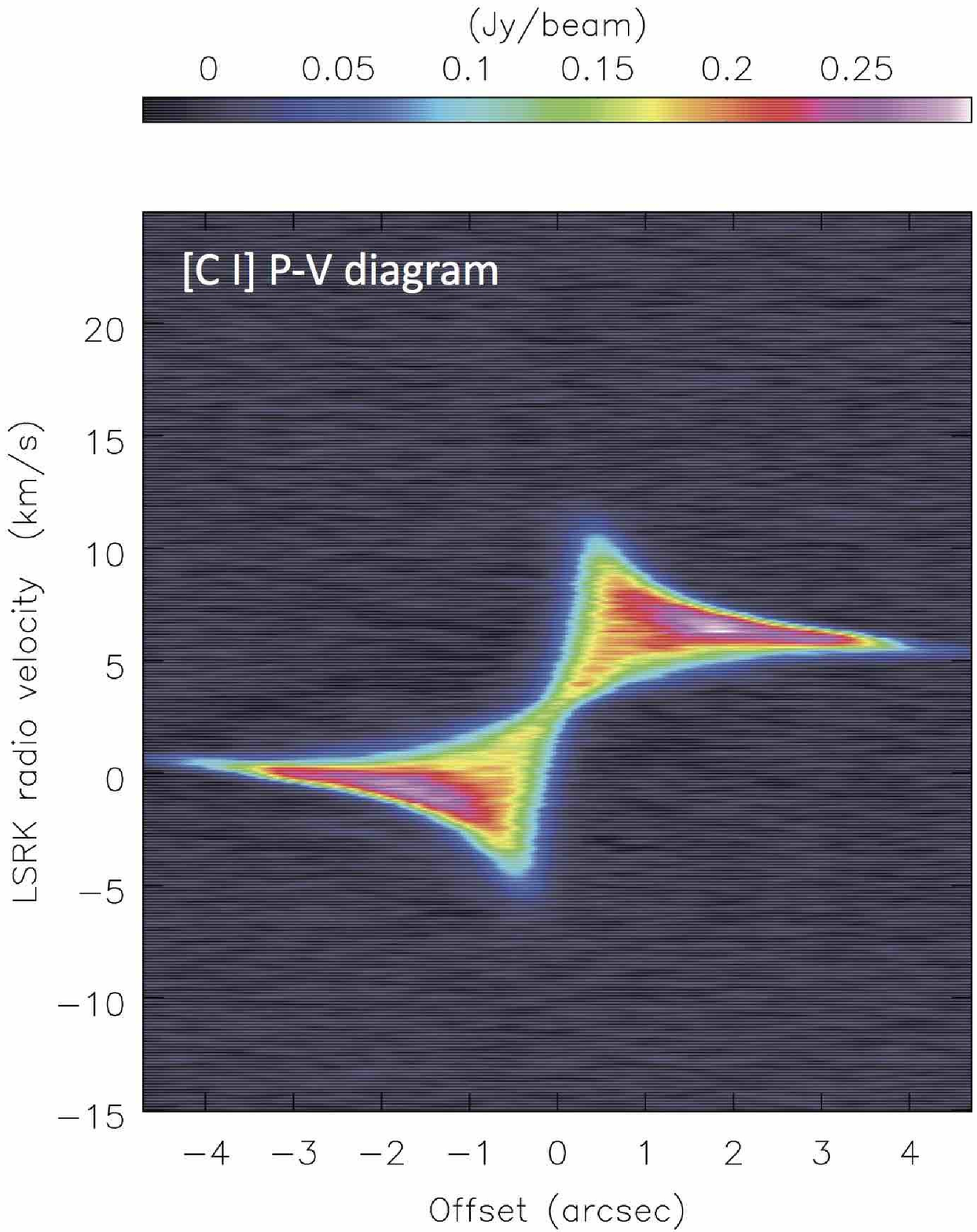}
\caption{The position-velocity diagram of [C I] emission along the major axis prepared by using the CASA {\it impv}. 
The starting position is ($\alpha_{J2000}$, $\delta_{J2000}$) = 
(01$^{\mbox h}$34$^{\mbox m}$38$^{\mbox s}.$210, ~$-$~$\!\!$15$^{\circ}$40$^{\arcmin}$36$\arcsec.$450),
while the end position is ($\alpha_{J2000}$, $\delta_{J2000}$) = 
(01$^{\mbox h}$34$^{\mbox m}$37$^{\mbox s}.$594, ~$-$~$\!\!$15$^{\circ}$40$^{\arcmin}$33$\arcsec.$450)
with P.A.= $-$ 72$^{\circ}$. The starting and end positions of the major axis are indicated in Figure \ref{fig1}(b).}
\label{fig4}
\end{figure}

\begin{figure}
\epsscale{1}
\plotone{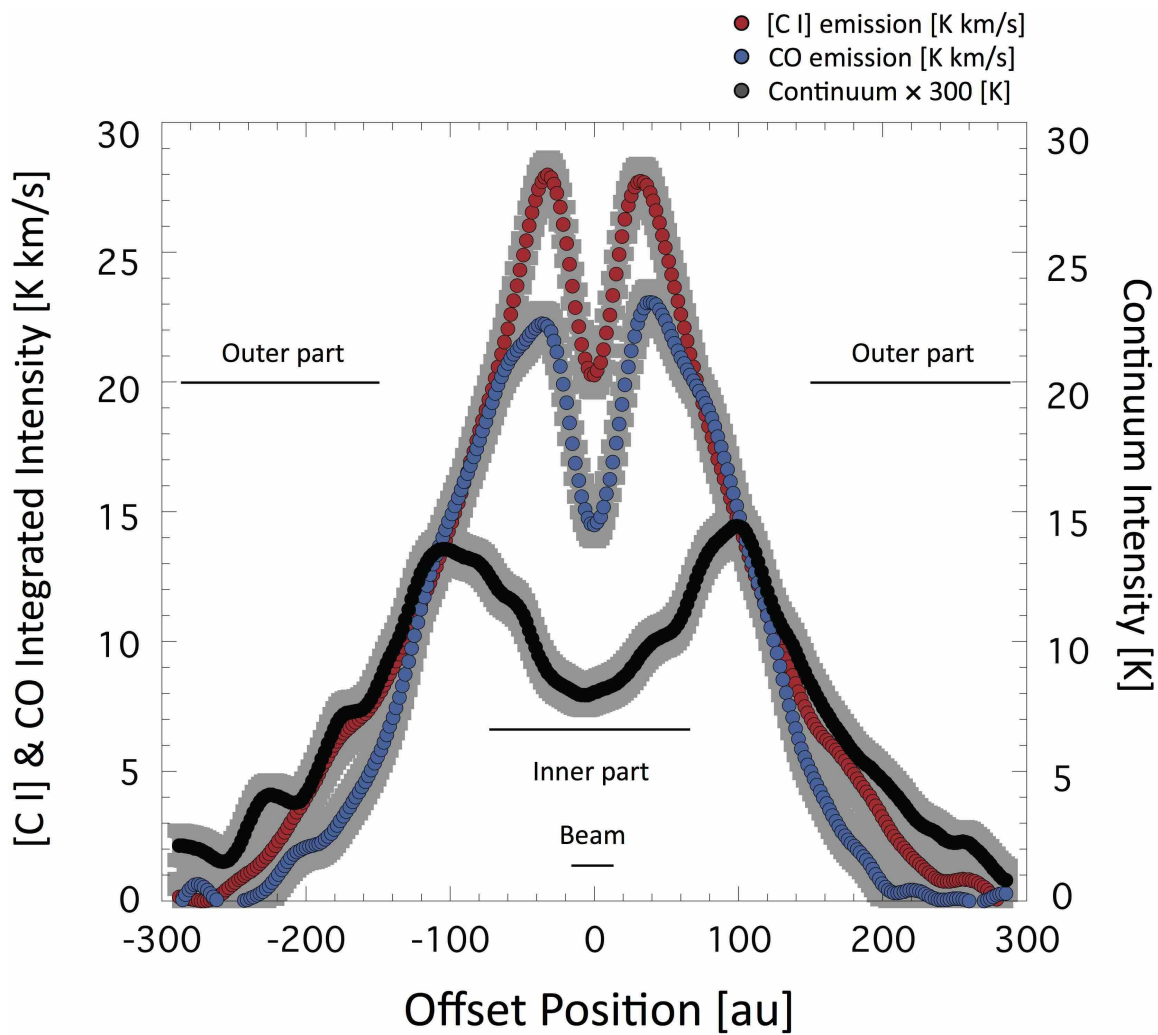}
\caption{The intensity of the [C I] (Red), CO (Blue), and 614~$\micron$ continuum (Black) emission as a function of the projected distance.
The gray shaded areas show the error bar that indicates the propagation of the 1~$\sigma$ rms noise level of the integrated intensity map.}
\label{fig2}
\end{figure}

\begin{figure}
\epsscale{1}
\plotone{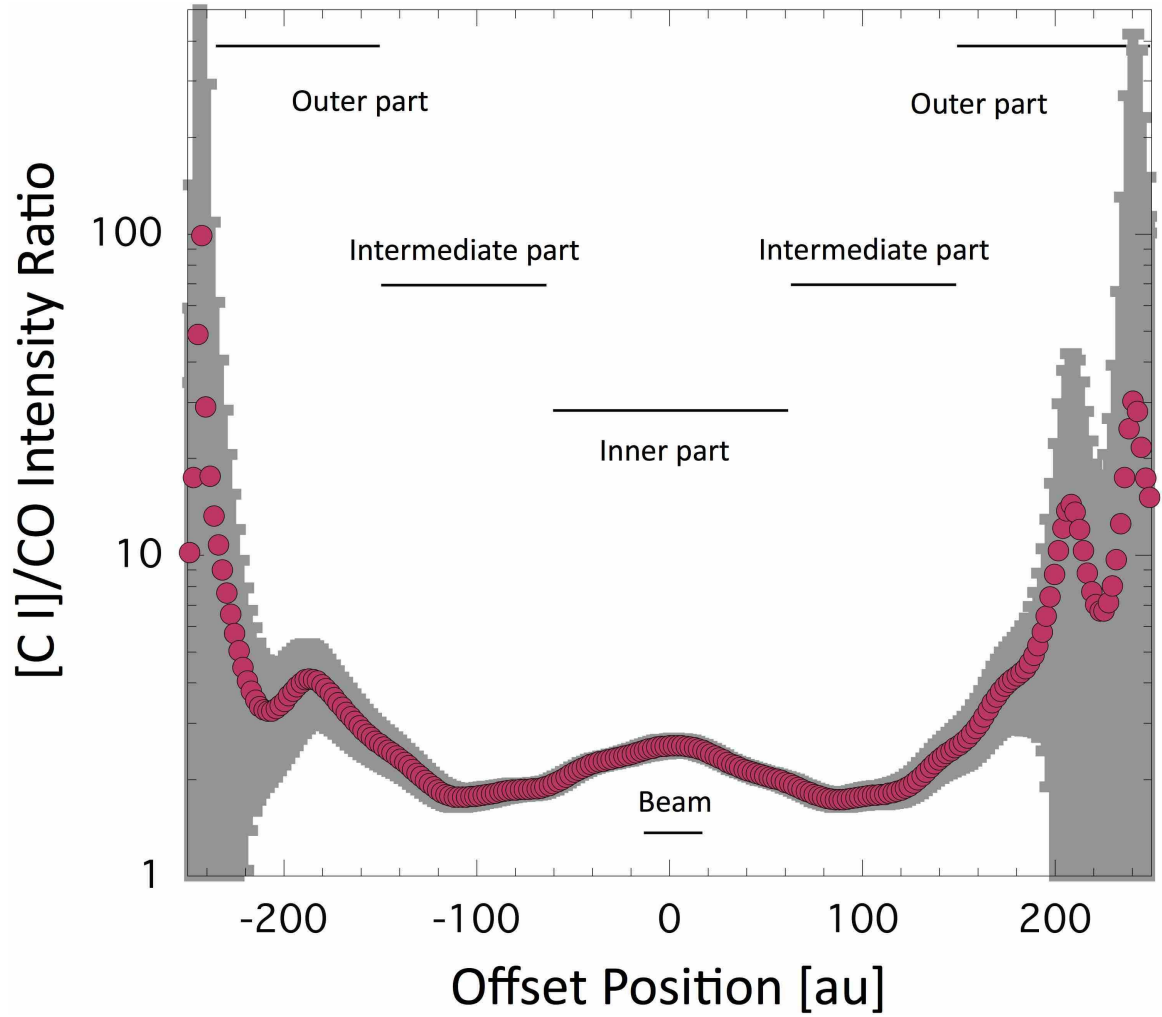}
\caption{The [C I]/CO intensity ratio as a function of the projected distance. 
The gray shaded area shows the error bar that indicates the propagation of the 1~$\sigma$ rms noise level of the integrated intensity map.
Since the error propagation employed here assumes the small errors of the CO and [C I] intensity, the error in the outer part may be underestimated.}
\label{fig3}
\end{figure}

\begin{figure}
\epsscale{1}
\plotone{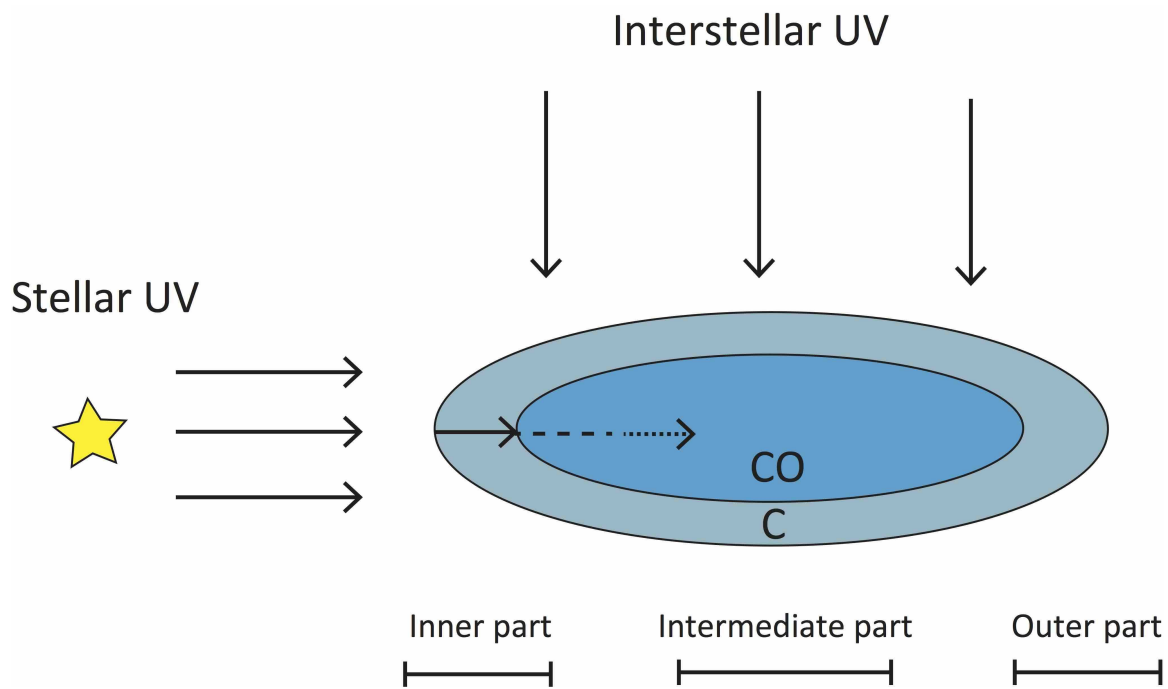}
\caption{The schematic picture of the vertical cross section of the broad ring structure for the interpretation of the variation of the 
[C I]/CO intensity ratio for the optically thick case (see Section 3.4).
We assume that CO and C are vertically segregated.}
\label{fig4}
\end{figure}

\clearpage
\appendix

\section{Size Estimation}

\begin{figure}
\epsscale{1.2}
\plotone{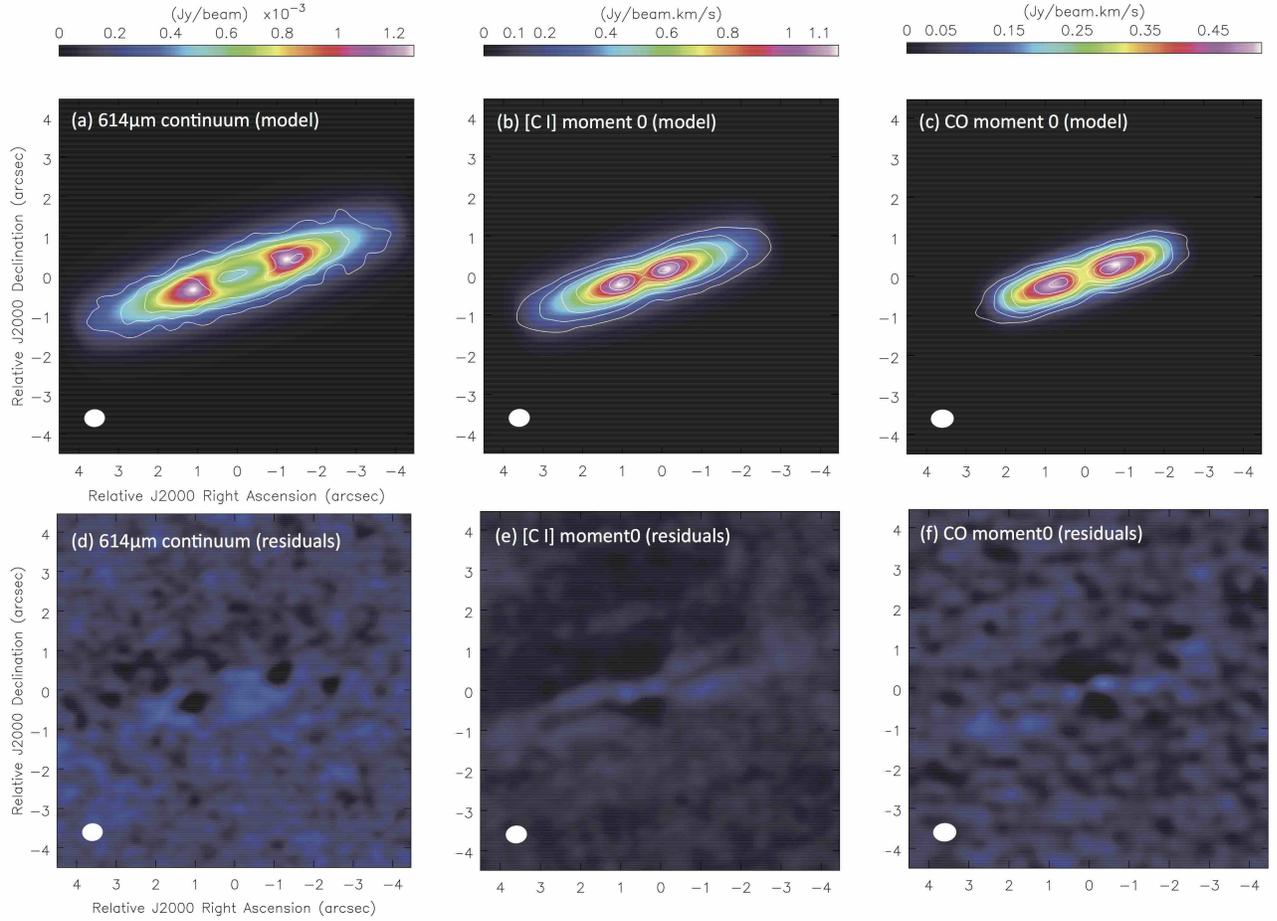}
\caption{
The best fit model images and their residuals.
(a) The 614~$\micron$ dust continuum model image (color) compared with the data (contours). 
Contours are 5$\sigma$, 10$\sigma$, 15$\sigma$, 20$\sigma$, 25$\sigma$ levels (1$\sigma$ = 0.045~mJy~beam$^{-1}$). 
(b) The [C I] integrated intensity map of the model compared with the data (contours). % (Velocity range = --5 to 11~km~s$^{-1}$).
Contours are 5$\sigma$, 10$\sigma$, 15$\sigma$, 20$\sigma$, 25$\sigma$, 30$\sigma$, 35$\sigma$, 40$\sigma$ levels 
(1$\sigma$ = 30~mJy~beam$^{-1}$~km~s$^{-1}$).
(c) The CO integrated intensity map of the model compared with the data (contours). % (Velocity range = --5 to 11~km~s$^{-1}$). }
Contours are 5$\sigma$, 10$\sigma$, 15$\sigma$, 20$\sigma$, 25$\sigma$, 30$\sigma$, 35$\sigma$, 40$\sigma$ levels 
(1$\sigma$ = 15~mJy~beam$^{-1}$~km~s$^{-1}$).
(d) Residuals map of the 614~$\micron$ dust continuum. 
(e) Residuals map of the [C I] integrated intensity.
(f) Residuals map of the CO integrated intensity.}
\label{figA}
\end{figure}

\begin{figure}
\epsscale{0.45}
\plotone{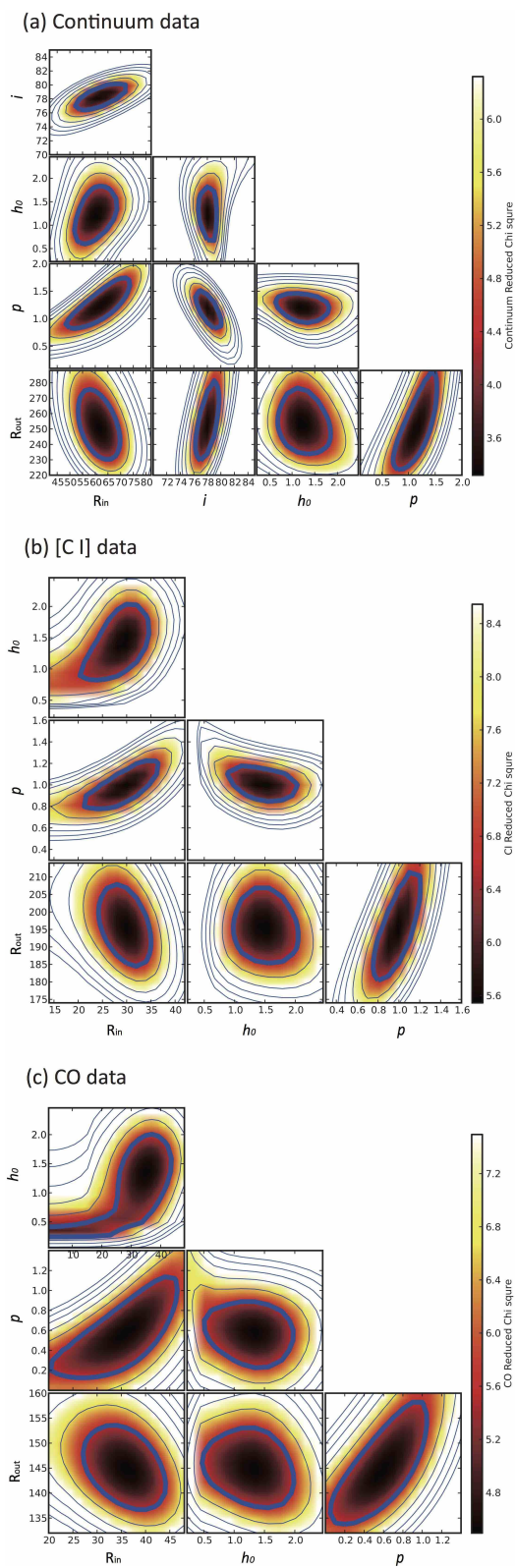}
\caption{Posterior distributions of $\chi^{2}$ for selected parameters of the model for (a) continuum, (b) [C I] and (c) CO data.
Contours show $\Delta{\chi}^{2}$ of 1, 2, 3, 4, 5, and 6. 
Boldface contours show $\Delta{\chi}^{2}$ of 1, which correspond to the 1$\sigma$ error (the 68$\%$ confidence level).}
\label{figB}
\end{figure}

In order to estimate the overall extent of the disk of the [C I], CO and dust continuum emission, 
a cut-off power-law disk model is created by using the equations (1) and (2) in \cite{hug17}. 
Assuming the optically thin condition, we use the spectrally integrated intensity in the analysis.
\cite{hug17} reported that their exponential model better fits than the power law model for the CO distribution.
In addition, the optically thin condition is not satisfied in our case (Section 3.4.).
However, our purpose of this analysis is not to model the disk structure in detail, but to just determine the approximate sizes of the dust, 
[C I] and CO disks.
Therefore, we employ the following equation of the surface flux density profile as a function of the radius ($R$), 
which is normalized to a radius of 100 au.

\begin{equation}
\sigma(R) = \sigma_{\rm{100au}} \left( \frac{R}{100~\rm{au}} \right)^{-p},
\end{equation}
where
\begin{equation}
\sigma_{\rm{100au}} = \frac{F_{\rm{tot}}\rm{(2-\it{p})}}{2\pi(100~\rm{au})^{\it{p}}({R_{\rm{out}}}^{2-\it{p}}-{R_{\rm{in}}}^{2-\it{p}})}.
\end{equation}
Here $F_{\rm{tot}}$ is the total flux (dust continuum: 0.03~Jy, [C I]: 17.2~Jy~km~s$^{-1}$, CO: 6.2~Jy~km~s$^{-1}$, estimated above the 3$\sigma$ emission of observed images), 
$p$ is the surface flux density power-law index, $R_{\rm{in}}$ is the inner radius, and $R_{\rm{out}}$ is the outer cut-off radius.
The radiation source density (in a unit of Jy~cm$^{-3}$) is calculated as a function of the radius and the height above the midplane $z$ as:
\begin{equation}
s(R, z) = \frac{\sigma(R)}{\sqrt{\pi}h(R)}\exp{\left(-\frac{z}{h(R)}\right)},
\end{equation}
where $z$ is the midplane of the disk.
$h(R)$ is the scale height
\begin{equation}
h(R)=h_{0}\sqrt{\frac{2{k_{\rm{B}}T(R)R^{3}}}{{m_{0}GM_{*}}}},
\end{equation}
where $M_{*}$ is the stellar mass, $m_{0}$(=2.33$m_{\rm{H}}$) is mass with a mean molecular weight, 
where $m_{\rm{H}}$ is the mass of the hydrogen atom, and $k_{\rm{B}}$ is the Boltzmann constant. 
As in \cite{hug17}, the temperature structure is assumed as:
\begin{equation}
T(R) = 40\left(\frac{R}{100~\rm{au}}\right)^{-0.5}.
\end{equation}
The best-fit result is listed in Table 1 and the corresponding model images and residuals are presented in Figure \ref{figA}.
A grid-based search is conducted to find the minimum $\chi^{2}$ (Figure \ref{figB}). 
This best fit result in the reduced $\chi^{2}$ is calculated to be 3.32 for the dust continuum, 5.55 for [C I], and 4.50 for CO, respectively.
As shown in the residual maps (Figure \ref{figA}), the model images capture major features of the distributions and hence, the model parameters are used for comparison of the distributions in Sections 3.1 and 3.2.

\begin{deluxetable}{l l l l l l l c c c c c c}
\tabletypesize{\small}
\label{tb1}
%\rotate
\tablecaption{Best-fit parameters}
\tablewidth{0pt}
\tablehead{
\colhead{Parameter} & \colhead{614~$\micron$ } & \colhead{[C I]} & \colhead{CO} \\
& \colhead{dust continuum} & \colhead{} & \colhead{}}
\startdata
%$M_{*}$ [$\MO$] & 2.1 & 2.1 & 2.1 \\
$R_{\rm{in}}$ [au] & 60 $\pm$ 10  & 30 $\pm$ 5 & 35 $\pm$ 10 \\
$R_{\rm{out}}$ [au] & 250 $\pm$ 20 & 195 $\pm$ 20 & 145 $\pm$ 10 \\
$p$ & 1.2 $\pm$ 0.3 & 1.0 $\pm$ 0.2  & 0.6 $\pm$ 0.4 \\
$h_{0}$ & 1.2 $\pm$ 0.4 & 1.4 $\pm$ 0.6 & 1.3 $\pm$ 0.7 \\
$i$ [$^{\circ}$] & 78 $\pm$ 2 & 78 $\pm$ 2 & 78 $\pm$ 2 \\
\hline
\enddata
\tablecomments{The single power-law model by \cite{hug17} is applied to our 614~$\micron$ dust continuum, [C I] and CO data.
The stellar mass of 2.1~$\MO$ is taken from \cite{hug17}. The error range is shown in the boldface contours in Figure \ref{figB}.}
\end{deluxetable}

\end{document}